\documentclass[twocolumn]{aa}  
\usepackage{txfonts}
\usepackage{graphicx}
\usepackage[]{natbib}
\bibpunct{(}{)}{;}{a}{}{,}

\begin{document}

\title{Are magnetic hot stars intrinsic X-ray sources ?}

\author{S. Czesla \and J.H.H.M. Schmitt}  
\institute{Hamburger Sternwarte, Universit\"at Hamburg, Gojenbergsweg 112, 21029 Hamburg, Germany}
\date{Received ... / accepted ...}

\abstract
{
X-ray surveys carried out with the \textit{Einstein}  and ROSAT satellites have resulted in  rather unexpected detections of
X-ray emission from late B-type and early A-type stars. These stars possess neither winds like early-type stars nor convective envelopes as late-type stars, 
so that the origin and production mechanism of this X-emission is unclear.
}
{
We investigate whether the presence of large magnetic fields is related to the observed X-ray emission.
}
{
We carried out \textit{Chandra} high-angular resolution observations of a sample of late B-type 
and A-type stars with measured magnetic fields in the range from $0.2 - 17$~kG. Out of 
the selected $10$ sample stars, $6$ objects had been previously detected as X-ray sources, some of them, however, with high positional uncertainty and a low signal-to-noise ratio, while $4$ of our sample
stars do have large magnetic fields but no previous detections of X-ray emission.
}
{
Our \textit{Chandra} data confirm all previous ROSAT detections with an extremely high significance, and the limits 
of the offsets between X-ray and optical positions are greatly improved. In 
particular, \object{HD 215441}, known as Babcock's star with the strongest magnetic field by far (17 kG) of our sample stars, 
a rather faint and somewhat marginal ROSAT source, can clearly be detected.
However, none of the $4$ ROSAT non-detections could be detected with the new \textit{Chandra} observations.
}
{
The pure existence of a magnetic field of kiloGauss strength on a late B-type or A-type star is 
therefore not necessarily a prerequisite for finding X-ray emission among these stars. 
Understanding the observed X-ray emission
from Babcock's star is a challenge for observational and theoretical astrophysics.
}

\maketitle

 \section{Introduction}
  
  One of the fundamental discoveries in stellar X-ray astronomy is
  that X-ray emission is found for almost all stars located 
  on the main sequence.   Early-type stars emit X-rays in proportion
  to their bolometric luminosities, and the X-ray emission is thought
  to be due to instabilities in these stars' radiatively-driven winds.
  Cool stars with outer convective envelopes produce X-rays following
  the rotation-activity paradigm, with the ultimate cause of the observed
  X-ray emission thought to be a magnetic dynamo operating in the
  interiors of these stars.   According to this standard paradigm,  
  X-ray emission is not predicted for A-type and late B-type stars, since
  these stars neither possess convection zones, which support the
  turbulent motions required for dynamo action, nor do they have
  strong winds to produce the shocks needed to explain the 
  X-ray emission observed from early-type stars.   
  
  Despite these expectations, X-ray emission has been found from late B-type and
  A-type stars. For example, within the framework of the ROSAT all-sky
  survey a fraction of $\approx$ 15 \% of all bright ($m_v <$ 6.5) A-type
  stars can be identified with a soft X-ray source.  A popular
  hypothesis to explain this ``unexpected" X-ray emission is to invoke
  unresolved faint companions as the source of the observed X-ray emission.
  These faint companions can be either white dwarfs, as in the 
  case of the well-known binary Sirius AB or the less well-known binary $\beta$ Crt
  \citep{fleming_1991}, or low-mass late-type companions or even post T Tauri stars, as in the large majority of the
  cases.
  Already \citet{golub_1983} pointed out that A-type stars are fairly
  young by necessity, thus causing any late-type companions to be quite
  X-ray luminous.
  
  Since A-type stars are rather bright optically, it is easy to
  ``hide" low mass stars in their vicinity.  Yet
  in a number of cases the ``binary hypothesis" can be tested directly.
  For example, for the totally eclipsing binary $\alpha$
  CrB (consisting of an A0 and a G5V type star), \citet{schmitt_kuerster_1993}
  were the first to discover a total X-ray eclipse during the
  optical secondary minimum of the system, thus demonstrating that essentially
  all of the system's X-ray flux does indeed come from the late-type 
  companion, as expected in the context of the ``binary hypothesis".
  Infrared high-resolution imaging observations of B-type stars detected 
  as X-ray sources in the ROSAT all-sky survey yielded a high fraction of new 
  and close companions.
  X-ray emission from such companions was searched for by
  \citet{stelzer_2006} in a sample
  of $11$ X-ray bright late B-type stars with known, close companions, but was previously 
  spatially-unresolved at X-ray wavelengths. \citet{stelzer_2006} did
  detect X-ray emission from  the low-mass companion in $10$ out of $11$ cases;
  however, in a surprisingly large number of cases  ($7$ out of $11$), X-ray
  emission was also found at the position of the primary late B-type star with 
  X-ray luminosity levels of L$_x\approx 10^{29}-2\cdot 10^{30}$~erg/s.  Therefore,
  these primaries must either be intrinsic X-ray sources or, alternatively, members
  of at least triple or higher multiplicity systems.  While a 
  fraction of $\approx 40$\% of the high multiplicity systems among X-ray detected 
  late B-type stars appears high, \citet{stelzer_2006} argue that 
  multiplicity studies among known spectroscopic binary stars of HgMn 
  type \citep[e.g. ][]{isobe_1991}
  reveal a compatible fraction of high multiplicity systems.
  Similar conclusions have previously been arrived at by \citet{schmitt_zinn_1993} using
  ROSAT HRI data.  Finally, \citet{kouwenhoven_2005} studied multiplicity among late B-type     and A-type \textit{Hipparcos}
  members of the Sco OB2 association -- containing $4$ of our sample objects -- and
  showed that at least $40-50$\% of the observed
  stars are members of multiple systems. \citet{kouwenhoven_2005} conclude that
  at least $41$\% of all \textit{Hipparcos} members of the Sco OB2 association show
  multiplicity and that this fraction still constitutes a lower limit due to
  observational biases. 
  
  In recent years, magnetic field detections have been obtained for 
  more and more early-type stars.  Also, X-ray emission phenomena
  have been connected to the presence of magnetic fields in early-type stars.  
  For example, \citet{gagne_1997} find a dependence of the X-ray flux
  of the oblique magnetic rotator \mbox{$\theta$ Ori C} on rotation, and flaring has been
  observed for $\lambda$~Eri \citep{smith_myran_1993} and the magnetic star
  $\sigma$ Ori E \citep{groote_schmitt_2004}.  The \textit{rigidly rotating
  magnetosphere model}, developed by \citet{townsend_owocki_2005} and 
  applied to $\sigma$~Ori~E \citep{townsend_owocki_groote_2005}, provides a
  possible explanation for the observed strength and variability of the
  observed quiescence radiation.  Furthermore, the simulations carried out by
  \citet{doula_townsend_2006} may provide an explanation of the flaring in the context
  of the same theoretical framework without the involvement of a late-type companion.
  Of course, the interpretation
  of the flares as magnetic phenomena associated with the early type star 
  is a controversial issue \citep{sanz_forcada_2004}.
  Finally, \citet{schmitt_2004} report a weak, and somewhat marginal,              
  (2 $\sigma$) detection of X-ray emission from Babcock's star
  (HD~215441), the main sequence star with the largest known magnetic
  field.
  Therefore, the question arises as to what extent the presence of a
  magnetic field is a prerequisite for the detection of X-ray emission
  from stars without convective envelopes and without strong winds.         
  
  In this paper we report on the \textit{Chandra} observations of a sample of
  B- and A-type stars with large magnetic fields.  Our paper is structured as follows.
  In Sect.~\ref{sec:Observations} we provide basic 
  information on our target stars and the new \textit{Chandra} observations; 
  we also describe our data analysis, and first results derived from
  the new X-ray data. In Sect.~\ref{sec:Individual} spectral analysis 
  and further properties of our sample objects are discussed.
  Finally, Sect.~\ref{sec:Discussion} and Sect.~\ref{sec:SummAndCon} present a
  discussion and a summary of our findings
  and put them into the context of our current understanding of stellar X-ray emission.
  In particular our results are discussed in view of binarity and the magnetically
  confined wind-shock model introduced by \citet{babel_montmerle_1997}.
  
 \section{Observations and data analysis}
  \label{sec:Observations}
  
  Our stellar sample comprises $10$ stars in the spectral range of B$8$ to A$6$. 
  All targets are known to possess strong magnetic fields as listed in the Bychkov catalog of 
  averaged stellar effective magnetic fields \citep{bychkov_2003} and were observed with \textit{Chandra}'s ACIS-S instrument with exposure times
  between $3$ and $20$ kiloseconds.  
  In Tables~\ref{tab:Targets1} and \ref{tab:Targets2} we list
  some basic information about the observed target stars and the {\it Chandra}
  observations.  Specifically we list the HD number, spectral type, and parallax of our target 
  stars, as well as sequence numbers and exposure times of our \textit{Chandra} 
  observations in Table~\ref{tab:Targets1}, while
  Table~\ref{tab:Targets2} lists more information about magnetic fields and previous 
  observations of X-ray emission with the ROSAT satellite.
  
  \begin{table}[h]
  \begin{minipage}{0.5\textwidth}
    \caption{Basic data on target stars and {\it Chandra} observations, with the
    Spectral type and parallax both taken from the SIMBAD data base.
    \label{tab:Targets1}} 
    \begin{tabular}{r r r r r r} \hline \hline
      HD & seq.num.\footnotemark[1] & sp.type &\multicolumn{2}{c}{$\pi$ [mas]} & obs.time [ks] \\ \hline
      12767  & 200326 & B9.5sp& $9.03 $&$ \pm 0.78$  & 3 \\
      15144  & 200327 & A6Vsp & $15.24$&$ \pm 0.95$ & 3 \\
      143473 & 200320 & B9    & $8.07 $&$\pm 0.99$  & 15 \\
      144334 & 200328 & B8V   & $6.7  $&$ \pm 0.86$ & 3 \\
      146001 & 200329 & B8V   & $7.06 $&$\pm 0.81 $ & 3 \\
      147010 & 200321 & B9II/III&$ 6.9$8&$\pm 0.95$ & 15 \\ 
      184905 & 200322 & A0p   & $6.06 $&$\pm 0.5  $ & 15 \\ 
      208095 & 200323 & B6IV-V& $5.07 $&$\pm 1.52 $ & 15 \\
      215441 & 200324 & A0p   & $1.4  $&$\pm 0.9  $ & 20 \\
      217833 & 200325 & B9IIIwe&$ 4.51$&$\pm 0.82 $ & 15 \\ \hline
    \end{tabular} \\[2mm]
    $^1$ the \textit{Chandra} sequence number
  \end{minipage}
  \end{table}
  
  \begin{table}[h]  
  \begin{minipage}{0.5\textwidth}
    \renewcommand{\footnoterule}{}  
    \caption{Data on magnetic field measurements, rotation velocities, and previous X-ray detections of the sample stars.
    \label{tab:Targets2}}
    \begin{tabular}{r| r r c r r l} \hline \hline
      HD     & v$\sin(i)$ & \multicolumn{3}{c}{magn.field [G]\footnote[1]{The magnetic field measurements, as well as the method used to obtain the field strength are taken from the Bychkov catalog of averaged stellar effective magnetic fields \citep{bychkov_2003}.}} & Method$^a$ & ROSAT \footnote[2]{A \mbox{$\surd$-sign} marks applicable properties. The paper showing the assignments to ROSAT X-ray sources is in preparation.}  \\ 
             & [km/s]     & \multicolumn{3}{c}{}               &        & det.                \\ \hline 
      12767  & 87         & 242.1 & $\pm$ & 93.9   & H1     & $\surd$ \\
      15144  & 14         & 802.5 & $\pm$ & 216.4  & Met    & $\surd$ \\
      143473 & -          & 4292.5& $\pm$ & 362.0  & all    & -  \\
             & -          & 4775.3& $\pm$ & 416.7  & Hl     &   \\
      144334 & 44         & 783.2 & $\pm$ & 257.7  & H1     & $\surd$ \\
      146001 & 181        & 647.2 & $\pm$ & 381.9  & Hl     & $\surd$ \\
      147010 & -          & 4032.1& $\pm$ & 402.7  & all    & -  \\
             & -          & 4050.7& $\pm$ & 466.5  &  Met   & \\
             & -          & 3594.4& $\pm$ & 379.8  &  Met   & \\
             & -          & 5096.0& $\pm$ & 324.0  & Hl     & \\
      184905 & -          & 5051.6& $\pm$ & 3039.4 & Met    & -  \\
      208095 & 120        & 7636.9& $\pm$ & 3206.3 & all    &   \\
      215441 & -          & 19437.3& $\pm$& 2086.5 & all    & $\surd$ \\
             & -          & 17543.2& $\pm$&  572.2 & Met    & \\
             & -          & 15555.9& $\pm$& 2032.1 & Hl     & \\
      217833 & -          & 3648.7 &$\pm$ & 697.5  &   all  & $\surd$ \\
             & -          &4105.5 &$\pm$  & 768.2  &  Met   & \\
             & -          & 748.9  &$\pm$ & 336.5  &   Hl   & \\ \hline
    \end{tabular} 
  \end{minipage}
  \end{table}

  \subsection{General procedures and X-ray detections}
   
   Analysis of the \textit{Chandra} X-ray data was carried out using the CIAO package (version $3.3$) 
   and is based on the photon event files provided by the \textit{Chandra} pipeline. As a first 
   step in the analysis, the event files were screened for photons with energies higher than
   $0.5$~keV to exclude low-energy background photons. Thermal X-ray spectra with typical coronal temperatures 
   provide most of the flux above the limit of $0.5$~keV.   We then examined
   those count events originating in the vicinity of the anticipated optical positions of
   our sample stars, which were taken from the SIMBAD data base 
   and appropriately corrected for proper motion.   Around the thus defined positions we extracted 
   all photons within a circular detect cell.  The detect cell sizes were calculated
   from the point response functions for \textit{Chandra}'s ACIS-S instrument as provided in
   the CIAO data base. We specifically used on-axis point response functions, with
   the energy of the contributing photons fixed to $1$~keV. From these simulations we
   deduced that $95$\% of the counts from a point-like source are contained in a circle with a
   $1.3$~arcsec radius. 
   This circle defines our detect cell, within which we counted all registered events in the
   energy band from $0.5-2$~keV. We excluded events with energies above $2$~keV here 
   (but not in the spectral analysis carried out in Sect.~\ref{sec:SpectralAnalysis}) to
   avoid losing sensitivity due to high-energy background.
   The effective background was estimated from
   nearby regions in the X-ray images in an annulus between $10$ to $40$~arcsec around the
   central position (containing no obvious X-ray sources) and scaled to the chosen
   detect cell size. In all of our observations the estimated background per detect cell
   was found to be less than unity.
   
   The results of this first step in the analysis of our {\it Chandra} data 
   are summarized in Table~\ref{tab:Results}.   As is clear from Table~\ref{tab:Results},
   clear detections were obtained
   for 6 out of 10 target stars; a comparison of the extracted
   detect cell counts with the expected detect cell backgrounds shows our detections to be
   extremely significant, so we refrain from a formal statistical analysis
   (the probability of misidentifying the target stars with spurious sources is well below $10^{-7}$).
   We mention in passing that in those cases we actually computed the center of gravity of
   the observed photon distribution, determined its sky coordinates, and computed
   the offset between X-ray and optical positions, which are also listed in
   Table~\ref{tab:Results}; and the extracted detect cell counts refer to the center of
   the observed photon distribution.  An inspection of the offsets shows most
   offsets to be small (i.e., $<$ 0.5 arcsec) and on the order of the pixel size;
   the two outliers \object{HD 146001} and \object{HD 217833} will be discussed below.
   
   We searched for short-term variability in the X-ray light curves and especially for flare signatures,
   but no significant variations were found.
   
   \begin{table}[h]
   \begin{minipage}{0.5\textwidth}
     \renewcommand{\footnoterule}{}  
     \centering
     \caption{Results from the analysis of the {\it Chandra} data. The table
     is separated into a detection section, and an upper limit section for the non-detections.
     \label{tab:Results}}
     \begin{tabular}{r r r r} \hline \hline
       HD\footnote[1]{See Sect.~\ref{sssec:HD15144} for the case of HD~$15144$.}     & offset\footnote[2]{The offset refers to the angular distance between X-ray and optical position.} & detect cell\footnote[3]{Detect cell as well as background counts listed in the table refer to the $0.5-2$~keV energy band.}  & expected bg.$^c$   \\
              & [as]     & counts  & counts                \\ \hline
       \multicolumn{4}{l}{Detections} \\ \hline
       12767  &  0.05    &  122    & 0.012        \\
       15144A &  0.25    &  359    & 0.014         \\
       15144B &          &   27    & 0.014        \\
       144334 &  0.28    &  354    & 0.04          \\
       146001 &  0.53    &   25    & 0.015       \\
       215441 &  0.13    &   72    & 0.081          \\
       217833 &  0.69    &   69    & 0.068        \\ \hline
       \multicolumn{4}{l}{Non detections} \\
              &          & source counts&                    \\
              &          & upper limits &  \\ \hline
       143473 &    -     & $<$2.9 &  0.248       \\
       147010 &    -     & $<$3.1 &  0.052        \\
       184905 &    -     & $<$3.1 &  0.049       \\
       208095 &    -     & $<$3.1 &  0.028       \\ \hline
     \end{tabular}
   \end{minipage}
   \end{table}

  \subsection{Non-detections and upper limits}
   
   Inspection of Table~\ref{tab:Results} shows that four of
   our sample stars (HD~143473, \object{HD 147010}, \object{HD 184905}, and \object{HD 208095}) were not detected
   in our {\it Chandra} data.  In the cases of HD~147010, HD~184905, and HD~208095, not a
   single photon was detected in the detect cell, and one event was detected in the detect
   cell around HD~143473; note that in these four cases the detect cells had to be placed
   on the nominal optical positions of the target stars rather than the 
   centroid of the best X-ray position.  The case of HD~143473
   is discussed further in Sect.~\ref{sec:HD143473}. For the other stars 
   we computed the expected source count, $\alpha_s$, yielding a measurement 
   of zero counts with a $5$\% probability assuming Poisson statistics.   
   Specifically we solved the equation  
   \begin{equation}
     \mbox{e}^{-0.95 \alpha_s - \alpha_b} = (1-0.95) \rightarrow \alpha_s=\frac{3-\alpha _b}{0.95} \; \frac{\mbox{counts}}{ \mbox{detect cell}}
     \label{eq:upper_limit}
   \end{equation}
   where the factor of $0.95$ accounts for $95$\% of the total counts contained
   in the detect cell, and $\alpha_b$ denotes the expected background count.  
   The resulting upper limit count rate is
   therefore given by $\alpha_s$/obs.time (see Table~\ref{tab:Results}).

 \section{Individual sources -- spectra and additional information}
 \label{sec:Individual}
  
  In this section we discuss the {\it Chandra} data of our target stars
  individually. Starting with a spectral analysis in Sect.~\ref{sec:SpectralAnalysis} 
  we proceed to give additional information on the
  properties of our sample objects in Sect.~\ref{sec:AdditionalInfo}.
  
  \subsection{Spectral analysis}
   \label{sec:SpectralAnalysis}
   
   We performed spectral analysis of our X-ray data for those sources with more than $50$~counts (cf. Table~\ref{tab:Results}).
   The methods applied to determining the spectral parameters are discussed in the following sections and the results
   summarized in Table~\ref{tab:FitResults}.
   \begin{table}[h]
   \begin{minipage}{\linewidth}
   \renewcommand{\footnoterule}{}  
     \centering
     \caption{Temperatures and X-ray luminosities derived from spectral fits or estimated from standard parameters.
     Errors given for the temperatures correspond to $1\cdot \sigma$ errors.
     \label{tab:FitResults}}
      \begin{tabular}[h]{r r r} \hline \hline
      HD     &   temperature                                  & L$_x$ \footnote[1]{\small refers to the (unabsorbed) $>0.5$~keV band}    \\
             &   [keV]                                        & [$10^{29}$~erg/s] \\ \hline
      \multicolumn{3}{c}{One temperature fits} \\ \hline
      12767  & $0.78^{+0.06}_{-0.06}$                         & $2.8$    \\
      15144A & $0.79^{+0.03}_{-0.03}$                         & $2.3$    \\
      215441 & $0.72^{+0.08}_{-0.12}$                         & $15$     \\ \hline
      \multicolumn{3}{c}{Two temperature fits} \\ \hline
      144334 & $0.85^{+0.12}_{-0.07}$ and $7.5^{+7.8}_{-2.3}$ & 31       \\ \hline
      \multicolumn{3}{c}{Luminosity estimates with standard assumptions} \\ \hline
      15144B & -                                              & $0.17$   \\
      146001 & -                                              & $1.2$    \\
      217833 & -                                              & $0.9$    \\ \hline
      \multicolumn{3}{c}{Upper limits for non-detections} \\ \hline
      143473 &    -                                           & $<$0.014 \\
      147010 &    -                                           & $<$0.02  \\
      184905 &    -                                           & $<$0.027 \\
      208095 &    -                                           & $<$0.04  \\ \hline     
      \end{tabular} 
   \end{minipage}
   \end{table}
   
   \subsubsection{HD~12767, HD~15144A, HD~215144}
    \label{sec:1TSpecs}
    The X-ray spectra of these stars can be fitted with an absorbed, one-temperature APEC model. Leaving all parameters --
    hydrogen column density, temperature, abundances, and normalization -- free, yields good fits  but poorly constrains
    estimates of hydrogen column density and abundances. Assuming a mean density of $1$~particle/cm$^3$ in the
    interstellar medium and fixing the hydrogen column density to the thus-obtained value does not lead to significantly worse fits. 
    Furthermore, fixing abundances to $0.3$ times solar abundances ($0.2$ for HD~12767) still results
    in good fits. The fitted temperature values are not sensitive to the choice of the other fit parameters, 
    and we note in 
    passing that different combinations of hydrogen column density 
    and abundance also result in good fits, so that these parameters remain basically undetermined.
    The obtained temperatures are listed
    in Table~\ref{tab:FitResults}. Errors correspond to $1\cdot \sigma$ values calculated with
    fixed column density and abundances. In Table~\ref{tab:FitResults} we also give the X-ray luminosity
    derived from the spectral fits.
   \subsubsection{HD~144334}
    \begin{figure}[h]
      \centering
      \includegraphics[scale=0.34]{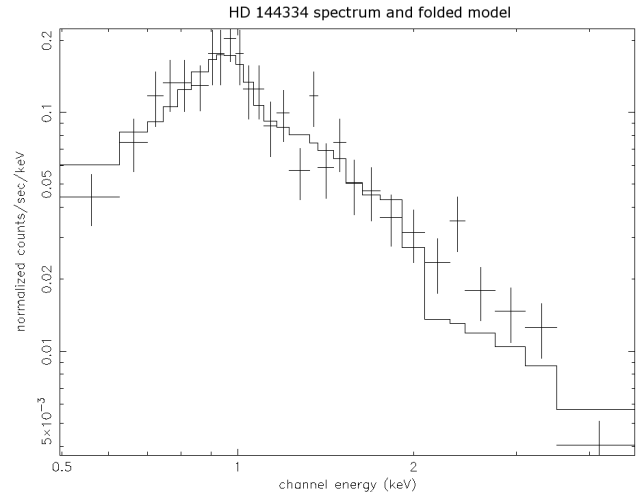}
      \caption{Spectrum and folded model for HD~144334. The spectrum is modeled using an absorbed, two-temperature APEC model.
      Hydrogen column density and abundances are fixed as described in the text. 
      \label{fig:144334Spec}}
    \end{figure}
    Figure~\ref{fig:144334Spec} shows the spectrum of HD~144334 and a model fit to the spectrum.  This spectrum cannot be fitted with 
    a one-temperature model, so we instead use an absorbed, 
    two-temperature APEC model to fit the data appropriately.  Unfortunately the data
    quality is such that - again - well-defined constraints cannot be derived for most parameters. 
    As in the cases discussed in Sect.~\ref{sec:1TSpecs}, fixed values for hydrogen column density, according to the assumption 
    of a mean interstellar density of $1$~particle/cm$^3$ and $0.3$ times solar abundance, result in a good fit (see
    Fig.~\ref{fig:144334Spec}). Table~\ref{tab:FitResults} lists the temperatures of both spectral components and the X-ray luminosity obtained 
    from the fit.
   \subsubsection{HD~15144B, HD~217833 and HD~146001}
    \label{sec:NoFit}
    Reasonable spectral fits cannot be obtained for these stars due to insufficient count statistics. Although more than $50$~counts
    were registered from HD~217833, no temperature could be derived. Thus, we used standard assumptions to estimate the
    X-ray luminosity of these objects. In particular, we used a hydrogen column density
    according to $1$~particle/cm$^3$ and a spectral energy distribution corresponding to a Raymond-Smith model at Log(T)=$7.0$
    and $0.4$ times solar abundances (compatible with the results of the spectral fits carried out for the other objects). 
    Applying these assumptions, we
    used the WebPIMMS tool to convert count rate into flux.
    The results are listed in Table~\ref{tab:FitResults}
   \subsubsection{Non-detections}
    For the non-detections (HD~$143473$, HD~$147010$, HD~$184095$, and HD~$209095$), we calculated an upper limit in L$_x$, by applying
    the standard assumptions outlined in Sect.~\ref{sec:NoFit} with the WebPIMMS tool.
    The upper limits are listed in Table~\ref{tab:FitResults}.
    
  \subsection{Comments on individual sources}
   \label{sec:AdditionalInfo}
   \subsubsection{HD 15144}
    \label{sssec:HD15144}
    \begin{figure}[h]
      \centering
      \includegraphics[scale=0.45]{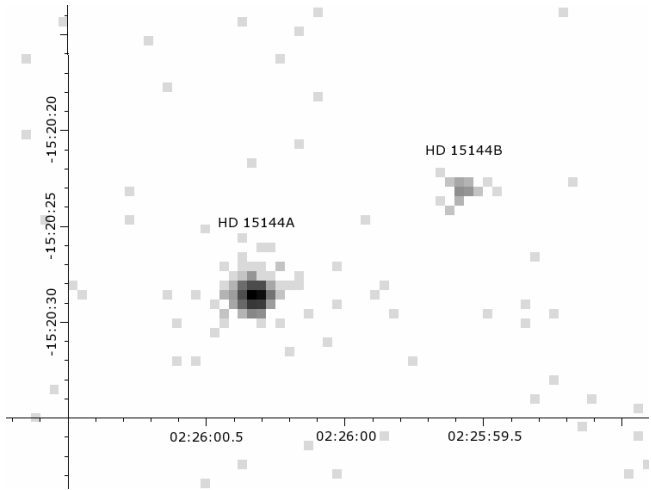}
      \caption{\textit{Chandra} image of the HD 15144 system.
      \label{fig:HD15144}}
    \end{figure}
    
    \noindent
    \object{HD 15144} is an A6V star with a magnetic field of $0.8$~kG.  With
    \textit{Hipparcos}, HD~15144 was resolved into two components, separated by $12.193$ arcsec at
    a position angle of $296^{\circ}$. Both components are clearly visible in our
    \textit{Chandra} image (see Fig.~\ref{fig:HD15144}).  The two X-ray sources
    are separated by $12.117$ arcsec at a position angle of $297^{\circ}$.  In
    Table~\ref{tab:Results} we therefore list both HD 15144A (the more luminous component both
    in the optical and at X-rays) and HD 15144B. From the magnitude difference measured
    by \textit{Hipparcos} we estimate a spectral type of $\sim$~G for the fainter companion, assuming
    it to be located on the main sequence. 
    Furthermore, \citet{leone_1999} point out that HD~15144A is actually a spectroscopic
    binary, the close companion not being resolved by \textit{Chandra}.  Unfortunately,
    \citet{leone_1999} do not discuss the nature of the companion, HD~15144Ab,  
    any further. They merely
    provide the mass function $f(m) = 1.81 \times 10^{-3} $ for this system.  Assuming 
    $i =  90^{\circ}$ and a mass of 2.0 M$_{\odot}$ for the primary, we derived a
    (minimum) mass of M $ \sim$ 0.2 M$_{\odot}$ for the secondary, and smaller inclinations
    lead to larger masses.   If we go on to assume that
    the orbit period of $2.99$~days is also the rotation period of HD~15144Ab, then HD~15144Ab
    is clearly a very plausible candidate for the observed X-ray emission at the position of
    HD~15144A.
    
   \subsubsection{HD~215441 - Babcock's star}
    
    Babcock's star (=HD~215441; spectral type A0p) is the star with the strongest magnetic field in our
    sample and in fact in the whole Bychkov catalog.  Our {\it Chandra} observations 
    (72 source counts) provide
    the first unambiguous evidence for X-ray emission from this unusual object;
    X-ray emission had been previously reported by \citet{schmitt_2004} on the basis of a rather faint ROSAT-PSPC source.
    Unfortunately the parallax of Babcock's star has a large error, the \textit{Hipparcos}
    catalog gives a value of $\pi=1.4\pm0.9$~mas, and consequently the computed X-ray 
    luminosity as given in Table~\ref{tab:FitResults} is also very uncertain.  No
    companion to Babcock's star is known.
    
   \subsubsection{HD~217833}
    
    The star HD~217833 is classified as B9III, and according to the
    \textit{Hipparcos} data it is a double system. 
    The angular separation of the optical components is given as
    $0.662$~arcsec at a position angle of 140$^{\circ}$.  We clearly detect HD~217833 
    in our \textit{Chandra} data (69 counts); however,
    the offset between the X-ray and optical positions is $0.69$~arcsec with
    a position angle of 145$^{\circ}$ (cf. Table~\ref{tab:Results}).
    Obviously the positional agreement is very much improved if we consider
    HD~217833B as the dominant contributor to the X-ray emission from the
    HD~217833 system. \textit{Hipparcos} does not reveal the nature of the companion but
    gives the magnitude difference between the primary and secondary components.
    From this we estimate a spectral type of $\sim$~F for the companion, assuming it to be 
    located on the main sequence. Due to the short lifetime of the primary B-star, the system must be
    young and the companion may still be in a pre-main sequence phase. In both cases the 
    companion is a plausible candidate for
    the observed X-ray emission. 
    
   \subsubsection{\object{HD 12767}, \object{HD 144334}, HD~146001}
    
    The stars HD~12767, HD~144334, and HD~146001 have spectral types B9.5, B8V, and B8V 
    respectively. All of these stars are clearly detected as X-ray sources with good 
    agreement between the optical and X-ray positions.  None of these stars is classified 
    as a binary, therefore  all are bona fide candidates for magnetic stars
    with X-ray emission.  We note that \textit{Hipparcos} indicates possible binarity for HD~144334, but an 
    adaptive optics near-infrared survey \citep{kouwenhoven_2005} shows no evidence of any stellar companion. \citet{kouwenhoven_2005} 
    also find no signatures of any companion around HD~146001 (cf. Table~\ref{tab:Binarity}).  
    Interestingly, HD~144334 and HD~12767 show optical variability. 
    HD~144334 is a variable of the SX Ari type with a period of $1.49$~days, HD~12767
    is an $\alpha^2$~Canum Venaticorum variable with a period of $1.89$~days, and both 
    SX Ari and $\alpha^2$~Canum Venaticorum type variables are rotational variables.
    The last refer to magnetic main sequence stars of spectral type \mbox{F$7$-B$8$} showing strong lines
    of silicon, strontium, chromium, and rare earths. The brightness and magnetic field both vary with
    rotational phase; amplitudes are $<0.2$~mag. The SX Ari type variables show basically the same 
    characteristics but are hotter 
    (spectral types B$0$-B$9$).
    As for HD~215441, the stars HD~12767, HD~144334, and HD~146001 
    are the best candidates for intrinsic X-ray emission from late B-type stars.
    
   \subsubsection{\object{HD~143473}, HD~147010, HD~184905, and HD~208095}
    \label{sec:HD143473}
    
    The stars HD~143473, HD~147010, HD~184905, and HD~208095 have spectral types
    B9, B9II/III, A0p, and B6IV-V, respectively.   No photons were recorded from
    HD~147010, HD~184905, and HD~208095, and a single event was found within the detect cell
    around HD~143473 (referring to the $0.5-2$~keV band).   Should this photon be attributed to HD~143473 and the star
    be considered an X-ray source?  We first note that, with an expected background 
    count of \mbox{$0.248$~counts/detect~cell}, the probability of finding exactly one background photon is $0.19$.  
    In the cases of HD~147010, HD~184905, and HD~208095, the probability of finding no background photon within
    the detect cell is about $0.95$.  
    Assuming no intrinsic X-ray emission from 
    these stars, the probability of recording no background photon in these four 
    detect cells is about $0.68$.
    
    \noindent
    We next note that the photon in question was 
    recorded at a distance
    of $0.9$~arcsec from the nominal source position.  According to our point 
    response modeling $\approx$ $10$\% of the source photons originating from a point 
    source should be located at distances $0.9$~arcsec or farther from the center; in other words, by assuming
    the photon in question is a source photon, the probability of recording this photon 
    at a distance of $0.9$~arcsec is $0.1$.  In conclusion, attributing the photon to the 
    background is the simplest and in fact more probable hypothesis, 
    and we therefore consider HD~143473 as a non-detection. 
    
    An infrared survey by \citet{kouwenhoven_2005} revealed no companions for HD~143473 and
    HD~147010. An object was found close ($6.09$~arcsec) to the position of HD~143473 but
    is classified as a background star.
    \citet{shatskii_1998} points out that 
    HD~208095 is a triple system. From the given (B-V) colors and assuming all components to
    be located on the main sequence, we estimate that both A and B
    components are of late B-type or early A-type ((B-V) $\approx 0$~mag), and the third component has a late spectral
    type ((B-V) $=2.53\pm 1.0$~mag). For HD~184905, at least
    \textit{Hipparcos} does not indicate binarity.
    
 \section{Comparison to ROSAT}
  
\begin{figure}[h]
    \centering
    \resizebox{\hsize}{!}{\includegraphics{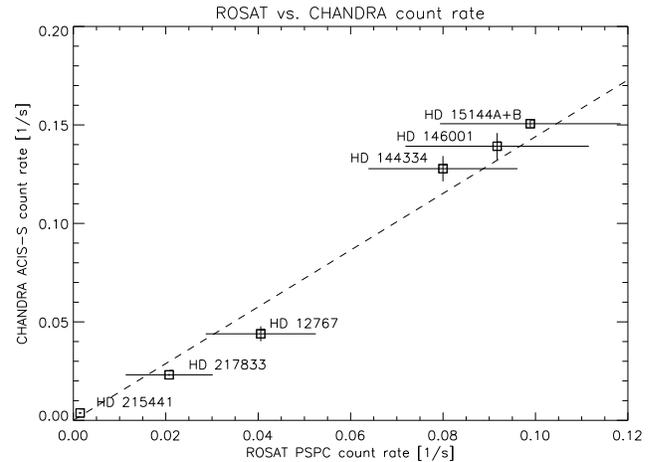}}
    \caption{\textit{Chandra} count rate vs. the ROSAT count rate; the dashed line is a linear regression. For HD~217833 and HD~146001,
    the \textit{Chandra} images show a second source close to the positions of the early-type components.  
    Since the ROSAT rates refer to the summed emission, we also
    used the summed \textit{Chandra} count rates for the correlation analysis.
    \label{fig:RoCha}}
  \end{figure}

  For $6$ of the stars listed in Table~\ref{tab:Targets2}, ROSAT detections and count rate measurements are available. 
  In order to test whether the X-ray properties of these objects have significantly changed between the ROSAT 
  and \textit{Chandra} observations, we
  plot (in Fig.~\ref{fig:RoCha}) the measured ROSAT-PSPC count rate                                                                 
  vs. the \textit{Chandra} count rate (referring to the $0.5-2$~keV band for \textit{Chandra}). 
  We apply the ROSAT count rates given in the $1$RXS catalog for all stars apart from HD~215441,
  which is listed only in the $2$RXP catalog
  (the paper presenting the assignment of the ROSAT sources is in preparation).  
  Obviously, the ROSAT and \textit{Chandra} count rates are correlated well and, given the
  substantial errors in some of the ROSAT count rates where most of these rates come from the
  all-sky survey, there appears to be no substantial variability between these X-ray observations that
  were taken typically more than ten years apart.
  
 \section{Discussion}   
  \label{sec:Discussion}
  
  \subsection{Binarity of the sample stars}
   \label{sec:Binarity}
   The information about binarity in our sample objects is summarized in Table~\ref{tab:Binarity}. Seven
   of our $10$ stars do not possess a {\bf known} companion and 
   are considered as {\it bona fide} single stars in the following;
   $1$ star belongs to a binary system, and another $2$ stars are members of 
   triple systems.
   Four of the $7$ single objects and $2$ of the multiple systems were detected as emitting X-rays.   
   \begin{table}
   \begin{minipage}{\linewidth}
     \renewcommand{\footnoterule}{}  
     \caption{Binarity and X-ray emission in the \textit{Chandra} sample. 
     In the last column
     we made the following assignments:
     H=\textit{Hipparcos}, CL=\citet{leone_1999}, K=\citet{kouwenhoven_2005}, and S=\citet{shatskii_1998}.
     \label{tab:Binarity}} 
     \begin{tabular}{r r r r r} \hline \hline
     HD     &  known            & single                        & X-ray      & data  \\
            &   components      & object\footnotemark[1]        & detection\footnotemark[1]  & origin     \\ \hline
     12767  &   1               & $\surd$       & $\surd$ & H \\
     15144  &   3               &  -            & $\surd$ & H\&CL \\
     144334 &   1               & $\surd$       & $\surd$ & K \\
     146001 &   1               & $\surd$       & $\surd$ & K \\
     215441 &   1               & $\surd$       & $\surd$ & H \\
     217833 &   2               & -             & $\surd$ & H \\
     143473 &   1               & $\surd$       & -       & K \\
     147010 &   1               & $\surd$       & -       & K \\
     184905 &   1               & $\surd$       & -       & H \\
     208095 &   3               & -             & -       & S \\
   \end{tabular} \\[2mm]
      $^1$ A \mbox{$\surd$-sign} marks applicable properties
   \end{minipage}
   \end{table}
   All of our sample stars that are also contained in the sample of \citet{kouwenhoven_2005} are classified
   as single stars. Two of these stars are detected in X-rays, while
   two are not. Assuming a K$_S$ magnitude of K$_S<12$~mag (corresponding to spectral type M$5$ or brighter) in the near infrared
   for possible companions and a magnitude difference $\bigtriangleup$K$_S<4.5$ for primary and secondary, 
   the angular separation between primary and any companion 
   cannot exceed $\approx 1$~arcsec; otherwise, the companion ought
   to show up in the survey. Note that
   \citet{kouwenhoven_2005} do detect companions with angular distances down to $0.22$~arcsec, 
   but this accuracy can only be reached under advantageous circumstances.  
   
  \subsection{Magnetic fields and X-ray luminosities}
  
\begin{figure}[h]
    \centering
    \resizebox{\hsize}{!}{\includegraphics{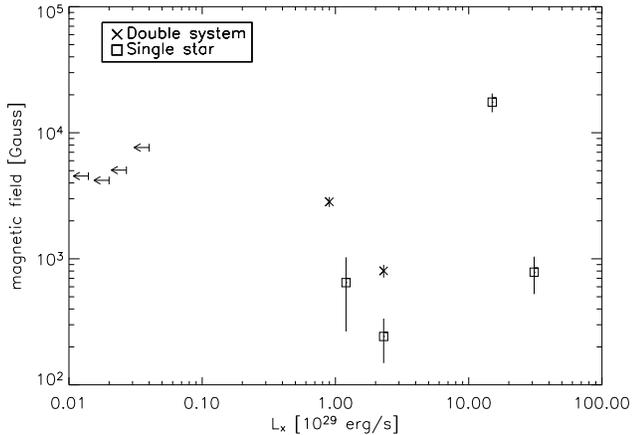}}
    \caption{Magnetic field vs. X-ray luminosity for sample stars; binary systems are marked with asterisks, squares indicate objects without
    known companions. The arrows mark the upper limits derived for non-detected stars. 
    \label{fig:LxVsB}}
  \end{figure}

  In Fig.~\ref{fig:LxVsB} we plot magnetic field strength $B$ against \mbox{X-ray} luminosity L$_x$. 
  The behavior of binary systems does not seem to differ from that of presumably single 
  stars so that a different nature of these objects is not required. The upper limits derived for the 
  non-detected stars are located more than an order of magnitude below the lowest 
  observed X-ray luminosity, suggesting
  substantially different X-ray properties for these objects.
  The behavior of the objects as illustrated in the plot is compatible with the companion
  hypothesis, assuming the non-detections to be single stars and the detected ones to be binaries with
  X-ray bright companions. The X-ray emission thus originates from otherwise unobserved late type or pre-main sequence
  companions. We note in passing that the cut off at $\approx 10^{29}$~erg/s below which no detection is found 
  represents the ROSAT detection limit. 
  Similar results are arrived at by \citet{daniel_2002} in an analysis of Pleiades member
  stars. Their results point to the existence two classes of late B to F-type stars: on the one hand, apparent X-ray
  emitters (L$_x>5\cdot 10^{28}$~erg/s) and, on the other, X-ray dark objects with upper X-ray luminosity limits $27-54$~times lower than
  the values obtained for the apparent emitters.  \citet{daniel_2002} also point out that indications 
  for the presence of a late-type
  companion exist in every case a late B to F-type star is found to be X-ray bright.
  
  \subsection{Application of the magnetically confined wind-shock model}
   
   In the following we demonstrate that the observed X-ray emission can
   also - alternatively - be interpreted in the frame of the 
   \textit{magnetically confined wind-shock model}.
   This model was introduced by \citet{babel_montmerle_1997} as
   a possibility to explain the X-ray emission from magnetic late B-type and A-type  stars,  and was applied to the magnetic A0 star IQ Aur.
   It predicts X-ray emission as the result 
   of the interplay of a radiatively driven stellar wind
   and the magnetic field; 
   the interaction occurs because sufficiently strong magnetic fields force the wind particles to move along 
   the magnetic field lines. In their calculations \citet{babel_montmerle_1997} assume
   a dipolar magnetic field structure, co-aligned with the rotation axis of IQ Aur, 
   limiting the regions of
   wind outflow to the surroundings of the polar caps. The equations
   of motion of the trapped particles were solved along the field lines.
   In the equatorial plane, the
   trapped wind components originating on either of the stellar hemispheres
   collide and shock. In the shock region, kinetic wind energy is 
   transformed and heats up the plasma, which subsequently
   radiatively cools through X-ray emission. \citet{babel_montmerle_1997} estimate 
   the shock temperature, $T_{sh}$, assuming a strong adiabatic shock, the specific heat
   ratio of $\gamma=5/3$, and a mean molecular weight of $\mu=0.5$ for the shocked material 
   from the Rankine-Hugoniot jump relation 
   \begin{equation}
     T_{sh}=1.13\cdot 10^5\mbox{ K} \left(\frac{v_w}{100\mbox{ km/s}} \right)^2 .
     \label{eq:Rankine}
   \end{equation}
   Here $v_w$ denotes the velocity of the shock-forming wind components. From their
   calculations \citet{babel_montmerle_1997}  derive a scaling law for
   the X-ray luminosity as a function of stellar magnetic field, $B_*$, wind velocity, $v_{\infty}$, 
   and a mass loss rate, $\dot{M}$, of the form 
   \begin{eqnarray}
     \mbox{L}_x &\approx& 2.6\cdot 10^{30} \frac{\mbox{erg}}{\mbox{s}} \left(\frac{B_*}{1\mbox{ kG}}\right)^{0.4} \xi \label{eq:ScalingLaw} \\
     \xi &=& \left(\frac{\dot{M}}{10^{-10} M_{\sun}/yr}\right)^{\delta} \left( \frac{v_{\infty}}{10^3\mbox{ km/s}} \right)^{\epsilon} , \nonumber
   \end{eqnarray}
   which makes
   the model applicable to a wide range of parameter values.   For $\delta$ and $\epsilon$,
   the values $\epsilon \approx 1-1.3$ and $\delta \approx 1$
   are derived from the calculations. As discussed by \citet{babel_montmerle_1997},
   the observed X-ray luminosity of IQ Aur necessitates very effective
   transformation of kinetic energy in the shock region to be compatible with
   the mass loss rate inferred from radiative wind models.
   
   Assuming then that at least some of the detected stars are indeed intrinsic X-ray emitters, we investigated to what extent the
   magnetically confined wind-shock model developed by \citet{babel_montmerle_1997} can be applied to interpret our data. We
   in particular assumed that the detected stars do possess radiatively driven winds and an approximately dipolar magnetic field configuration. 
   Applying Eq.~\ref{eq:Rankine}, we computed the
   wind velocities required to produce the temperatures indicated by our spectral analysis  (see Table~\ref{tab:FitResults}) and used  
   Eq.~\ref{eq:ScalingLaw} to derive the
   mass loss rates predicted by the model (we assume that $v_{\infty}=v_w$). 
   The results of our calculations are listed in Table~\ref{tab:BabMontResults}.
   For HD~144334 two parameter values are derived corresponding to the two spectral components
   needed to fit the spectrum -- we note that the characteristics of the hotter component 
   are only loosely determined. For IQ Aur, \citet{babel_montmerle_1997} determine wind
   velocities of $v_{eq}=500-900$~km/s at the equator (where the shock occurs) and a mass loss
   rate of $4.5\cdot 10^{-11}$~M$_{\sun}/$yr. These values are compatible with those derived
   for our objects.
   \begin{table}[h]
     \centering
     \caption{Wind velocity, $v_w$, and mass loss rate as derived from the application of
     the magnetically confined wind-shock model. For HD~144334, two values are listed for
     each parameter corresponding to two spectral components (cf. Table~\ref{tab:FitResults}).
     For the cool component, we derive L$_x=5.7\cdot 10^{29}$~erg/s and L$_x=2.55\cdot 10^{30}$~erg/s
     for the hot one.  
     \label{tab:BabMontResults}}
    \begin{tabular}{r r r} \hline \hline
    HD       &   $v_w$ [km/s]  & $\dot M$ [$10^{-10}$ $M_{\sun}$/yr]  \\ \hline
    12767    &   895           &  0.2  \\
    15144A   &   901           &  0.1  \\
    144334   &   934 / 2770    &  0.26 / 0.4 \\
    215441   &   860           &  0.2 
    \end{tabular}
   \end{table} 
   We, thus, conclude that the magnetically confined wind-shock model is a possible explanation for 
   the X-ray emission observed from our sample objects. 
   
 \section{Summary and conclusion}
  \label{sec:SummAndCon}
  
  In this paper we present \textit{Chandra} observations of $10$ intermediate-mass 
  stars with substantial magnetic
  fields between $0.2$ and $17$~kG.   Six out of our
  sample of $10$ stars show X-ray emission.   ROSAT detections are
  available for these stars although the ROSAT detection is somewhat marginal
  in the case of Babcock's star.
  Two of the detected stars in
  our sample, HD~15144 and HD~217833, are visual binaries and were
  resolved with \textit{Chandra} for the first time.  In both cases
  the companions are of late-type. For HD~217833, the
  binary separation is quite close ($0.6$~arcsec) and all of the 
  observed X-ray emission can be attributed to HD~217833B.  In the
  case of HD~15144 with a binary separation of 12~arcsec, two X-ray sources
  are clearly seen (cf. Fig~\ref{fig:HD15144}); however, the primary component itself is a spectroscopic binary
  with a possible late-type companion. 
  Significant short-term variability could not be detected for any of the sources.
  
  The remaining four stars (HD~12767, HD~144334, HD~146001, HD~215441 = Babcock's star)
  all possess large magnetic fields and 
  are clear X-ray sources with X-ray positions in
  excellent agreement with their optical position.  None of these
  stars is known to be binary.  Crude spectral analysis can be carried out
  for some of the detected sources, and it indicates X-ray temperatures 
  of $\approx$ 0.8 keV.  With these numbers it is then possible to
  demonstrate that the detected X-ray emission can in principle
  be explained in terms of the magnetically confined wind-shock model. 
  Therefore, a companion must not necessarily be held responsible for the X-ray emission. 
  
  However, our sample comprised $6$ stars with magnetic field strengths well above $1$~kG, 
  but only $2$ (HD~215441 and HD~217833, the latter a binary) were detected as X-ray sources, 
  suggesting that a large magnetic field strength need not necessarily lead to large X-ray emission
  and other parameters, such as wind speed, are more important.  Obviously, the sample size necessitates extreme caution and prevents
  any firm conclusion.  Clearly, in stars with such large magnetic fields as found, for example, in Babcock's star, we
  expect any convective motions to be efficiently suppressed.  Therefore a
  dynamo operation is rather hard to envisage.  The X-ray emission
  from Babcock's star can be interpreted in the framework of
  the confined wind-shock model, while
  in the binary scenario a companion would have
  to be held responsible for the detected X-ray emission.
  Therefore
  Babcock's star represents a true challenge to our understanding of X-ray emission
  from stars without convective envelopes.  It is certainly worthwhile carrying out deep searches for late-type companions, as well as searches
  for spectroscopic wind signatures.
  
  \bibliographystyle{aa}
  \bibliography{5890}
  
\end{document}